\documentclass[narrowdisplay]{elsart1p}
\usepackage{graphicx}
\usepackage{slashed}
\usepackage{amsmath}
\usepackage{lineno}

\def \bsgamma{b \to s \;\gamma}     %
\def \bsll{b \to s l^+l^-}     %

\newcommand{\be}{\begin{equation}}
\newcommand{\ee}{\end{equation}}
\newcommand{\bea}{\begin{eqnarray}}
\newcommand{\eea}{\end{eqnarray}}

\def\ba{\begin{array}}
\def\ea{\end{array}}
\def \bit{\begin{description}}
\def \eit{\end{description}}
\newcommand{\Mathematica}{\textsc{MATHEMATICA}}
\newcommand{\Maple}{\textsc{Maple}}
\newcommand{\MasterTwo}{\textsc{MasterTwo}}
\newcommand{\FeynArts}{\textsc{FeynArts}}
\newcommand{\Fermions}{\textsc{Fermions}}
\newcommand{\Integrals}{\textsc{Integrals}}
\def \eps {\epsilon}
\def \e {\epsilon}
\newcommand{\ga}{\gamma}

\newcommand{\li}{\rm Li}
\newcommand{\cl}{\rm Cl}
\newcommand{\oas} {${\cal O}(\alpha_s)$}

\newcommand{\ope}{{\cal O}}
\newcommand{\ty}{\texttt}
\def \nnl{\nonumber\\}    %
\newcommand\rmi {\ensuremath{i}}
\newcommand{\f}{\frac}
\newcommand{\gen}{\textbf{T}}
\newcounter{bla}

\begin{document}
\begin{frontmatter}

\title{
\MasterTwo: A Mathematica Package for the Automated Calculation of \\Two Loop Diagrams in the Standard Model and Beyond 
}

\author[a,b,c]{ S. Schilling\thanksref{author}},

\thanks[author]{Corresponding author: sabine.schilling@protonmail.com}

\address[a]{Institute for Theoretical Physics,
 University of Zurich,    \\
 Winterthurerstrasse 190,
 8057 Zurich, Switzerland}
\address[b]{Institute of Molecular Systems Biology,
ETH Zurich,
Wolfgang-Pauli-Str. 16,
8093 Zurich, Switzerland}
\address[c]{Lucerne School of Business,
Lucerne University of Applied Scienes and Arts,
R\"osslimatte 48,
6002 Lucerne, Switzerland}

\begin{abstract}
The calculation of rare loop decays in the Standard Model of Particle Physics and its extensions is an extremely tedious work. The \Mathematica~ package \MasterTwo\ facilitates this task. It automatically calculates all loop integrals reducible to scalar integrals depending on up to two different masses independent of external momenta. \MasterTwo\ consists of two sub packages, \Fermions~ and \Integrals. Whereas \Fermions\ covers the standard Dirac Algebra, \Integrals\ performs the Taylor expansion, partial fraction, tensor reduction and the integration of the thus achieved scalar integrals. The package works completely inside \Mathematica and can be easily customised for both educational and research purposes.


\end{abstract}

\end{frontmatter}

{\bf Package summary}\\
\begin{small}
\noindent
{\em Manuscript Title: \MasterTwo: A Mathematica Package for the Automated Calculation of \\Two Loop Diagrams in the Standard Model and Beyond 
}     \\
{\em Authors: S.Schilling} \\
{\em Package Title: \MasterTwo} \\
 {\em Version: 1.0}\\
 {\em Journal Reference:} \\
{\em Catalogue identifier:} \\
{\em Licensing provisions:} None \\
{\em Programming language:} \Mathematica \\
{\em Computer:} 
Computers running {\Mathematica} \\
{\em Operating system:} Linux, MacOs, Windows\\
{\em RAM:} Depending on the complexity of the problems 
{\em Classification:} Computer Algebra\\
{\em Nature of the physical problem: One- and two-loop integrals reducible to scalar integrals independent of external momenta and dependent on up to two different masses.}\\
{\em Solution method:} Heavy Mass Expansion and recurrence relations to transform tensor integrals to a larger number of scalar master integrals. Loop integration of the thus obtained scalar master integrals. \\
{\em Running time: Strongly depending on the problem and nature of diagram being calculated }\\

\end{small}
\tableofcontents
\section{Introduction}
The calculation of rare (loop) decays in the Standard Model of Particle Physics (SM) is an extremely complex, error prone task, which should be (semi) automated with the help of computer algebra programs. The \Mathematica\ package \MasterTwo\ was originally designed to facilitate such calculations arising in one and two-loop B-decays like $\bsgamma$ and $\bsll$ in the Standard Model of particle physics and and Two-Higgs-Doublet model extensions \cite{Schilling:2005cw}. \\
 \MasterTwo\ allows the automated calculation of all one- and two-loop integrals reducible to scalar integrals independent of external momenta and only dependent on up to two different masses. To do so it uses both the Heavy Mass Expansion \cite{Smirnov} and recurrence relations \cite{DT93} to transform tensor integrals to a larger number of scalar master integrals, which are then automatically integrated. The package's lean structure should make it an ideal candidate for both research and educational purposes. 
\MasterTwo\ consists of two sub packages, \Fermions~ and \Integrals. \Fermions~ contains all routines regarding the Dirac Algebra, \Integrals~ summarises all routines concerning the tensor reduction, partial fraction of one and two-loop integrals and the subsequent integration of the thus arising scalar integrals. This manual is organized as follows: 
The physics background and mathematical methods involved are summarized in sections \ref{sec:fermions}(package \Fermions) and \ref{sec:integrals} (package \Integrals), whereas sections \ref{subsec:docfermions} and \ref{subsec:docintegrals} document the corresponding \Mathematica\ commands. Section \ref{sec:installation} documents the installation of the package on the different operating systems Linux, Mac and Windows, whereas section \ref{sec:FeynArts} shows how the Standard Model output generated by the package \FeynArts\ can be adapted for the further usage in \MasterTwo. Finally we demonstrate the usage of \MasterTwo\
on an example diagram of the two-loop decay $\bsgamma$
in section \ref{sec:example}.

\section{\Fermions}
\label{sec:fermions}
\Fermions\ can simplify Dirac expressions in $D$ dimensions \
with an anticommuting $\gamma_5$. It provides the tools for standard
operations like contracting indices, sorting expressions and the use of the Dirac
equation. To calculate physical quantities as cross sections and decay-rates it allows furthermore to conjugate and square Dirac expressions and to compute traces over products of $\gamma$ matrices (for details see section \ref{sec:traces}).

\subsection{Declarations and Constants}
Before usage of the package, all arising masses, momenta, indices and polarisation vectors must first be declared. The documentation of the corresponding commands is given in \ref{sub:declarationsfermions}.\\

There are a few constants predefined in \Fermions:
\begin{itemize}
\item \ty{d} denotes the space-time dimension.\footnote {
 The {\bf{capital}} letter {\ty{D}}, which usually denotes space-time dimensions, is already used inside \Mathematica\ to indicate partial derivates. However, for reasons of better readability, $D$ will be used in all formulae of this manual to indicate the space time dimensions.}
\item \ty {eps} stands for $\epsilon$.
\item \ty{L} and \ty{R} are the left- and the right-projectors,
 respectively: $\ty{L}= 1/2 \;(1 - \gamma_5)$ and $\ty{R}= 1/2 \;(1 +
 \gamma_5)$.
\item \ty{Gamma5} stands for $\gamma_5$.
\item \ty{Unit} denotes the unit matrix.
\item \ty{Sigma[mu,nu]} is the tensor $\sigma_{\mu\nu} = \rmi /2 [\gamma_\mu,
\gamma_\nu]$.�\end{itemize}
The symbols \ty{L}, \ty{R} and \ty{Gamma5} are treated as projectors and, provided the expression is
simple enough, are shifted to the left automatically in order to reduce
the number of different terms. An expression like $\gamma_\mu L + R
\gamma_\mu$ will therefore automatically be transformed into $2 R \gamma_\mu$.

\subsection{Notation and Syntax}

After all the necessary declarations have been established, the corresponding
symbols can be used inside Dirac expressions and alike.

Gamma matrices, tensors and projectors like $\gamma_\mu,\;\sigma_{\mu\nu}, \mathtt{L}, \mathtt{R}$ and \ty{L} and {R} matrices are given as expressions with the head \ty{Dirac}, whereas scalar products are input as the function \ty{Scal}.
A few examples of simple structures:
 \begin{center}
 \begin{tabular}[h!]{ll}
 $g_{\mu\nu}$ & \ty{Scal[mu,nu]}, \\
 $p_\mu$ & \ty{Scal[p,mu]}, \\
 $p \cdot q $ & \ty{Scal[p,q]}, \\
 $ \textbf{{1}} $ & \ty{Dirac[]} (unit matrix in Dirac space),\\
 $ \gamma_\mu $ & \ty{Dirac[mu]}, \\
 $ \slashed{p} $ & \ty{Dirac[p]}, \\
 $ \gamma_5 $ & \ty{Dirac[Gamma5]} and similar for $L$ and $R$, \\
 $ \sigma_{\mu\nu} $ & \ty{Dirac[Sigma[mu,nu]]}.
 \end{tabular}
 \end{center}

\noindent

Some more complicated structures involving products of $\gamma$ matrices and
four-vectors might read:
\begin{center}
\begin{tabular}[h!]{ll}
 $ L \gamma_\mu \gamma_\nu \slashed{p} $ & \ty{Dirac[L, mu, nu, p]},
 \\
 $ p_\mu \gamma^\mu $ & \ty{Scal[p, mu] Dirac[mu]}, \\
 $ \gamma_\mu (\slashed{p} + m_b) \slashed{q}$ & \ty{Dirac[mu, p +
 mb, q]}, \\

 $ R (m_b \gamma_\mu + p_\mu) \gamma_\nu $ & \ty{Dirac[R, mb mu +
 Unit Scal[p, mu], nu]}.
 \end{tabular}
 \end{center}

\noindent
Note that masses inside Dirac structures need not to be provided
with an extra \ty{Unit} matrix, whereas this is indispensable for other structures like scalar products.

\Fermions\ makes no difference between covariant (up) and contravariant
(down) indices. It simply assumes that - if the same index appears twice - 
one is upper and the other lower and, if requested, takes the sum
over them.

\subsection{Dirac Algebra and Naive Dimensional Regularisation}
\label{sec:DiracAlgebra}
	The D-dimensional metric tensor $g$ is introduced satisfying
	\be
\label{metrictensor}
	g_{\mu\nu}g^{\nu \mu} =g_\mu^\mu=D,
	\ee
	where $D=4-2 \e $ in all kind of expressions containing Lorentz indices.
	The Dirac gamma matrices $\gamma^{\mu}=(\gamma^0, \gamma^i)$, where the Latin index $i$ is employed to denote spatial indices 1,2,3, satisfy the anticommutation relations
	\bea
	\label{anticommutation}
	\{\gamma^\mu, \gamma^\nu\}=2 g^{\mu \nu}&=&2 g_{\mu\nu}.
	\eea
	The $\gamma_5$ is defined by
	\be
	\label{gamma5}
	\gamma_5 =\gamma^5=i \gamma^0\gamma^1 \gamma^2 \gamma^3
	\ee
	and anti-commutes with all $\gamma^\mu$:
	\be
\label{commutation}
	\{\gamma^5, \gamma^\mu\}=0.
	\ee
It has been emphasised in the literature
that this rule leads to algebraic inconsistencies \cite{Breitenlohner:1975hg,Bonneau:1980ya}. Indeed, the naive dimensional regularisation (NDR) is inconsistent with
\be
\label{trcond}
Tr (\ga^\mu \ga^\nu \ga^\rho \ga^\sigma \ga_5)\neq 0
\ee
for dimensions of space-time $D=4-2\e$, $\e \neq 0$ .
However the latter condition is often considered to be necessary for
an acceptable regularisation, since at $D=4$ we must find
\be
\label{epsten}
Tr (\ga^\mu \ga ^\nu \ga^\rho \ga^\sigma \ga_5) = 4 \mathrm{i} \varepsilon^{\mu \nu \rho \sigma} \;.
\ee
Provided one can avoid the calculation of traces like eq.~(\ref{epsten}) containing $\gamma_5$ matrices, it has been demonstrated in many explicit calculations \cite{Buras:1989xd} that the NDR gives correct results consistent with schemes without the $\gamma_5$ problem.
From eq.~(\ref{gamma5}) we get for the projectors $R=(1+\gamma^5)/2$ and $ L=(1-\gamma^5)/2$
 \be
 \label{projector}
 \gamma^0 R = L\gamma^0, \; \gamma^0 L= R\gamma^0.
 \ee
The package does not need an explicit representation of the algebra, it can thus handle objects of the form $\gamma_\mu \gamma_\nu$ rather than
e.g. $\gamma_0 \gamma_2$. The function \ty{DiracAlgebra} performs the standard Dirac algebra according to eqs.~(\ref{metrictensor}), (\ref{anticommutation}), (\ref{commutation}) and (\ref{projector}). Further functions of \Fermions\ (conjugations, traces) are documented in section \ref{subsec:docfermions}.

\section{\Integrals}
\label{sec:integrals}
\Integrals\ performs all the steps necessary to transform the integrals into scalar master integrals of up to two different masses and independent of external momenta and their subsequent loop integration. A full list of all the available commands is given by the command \ty{IntegralsInfo[]}. A detailed documentation of all the functions introduced below is given in section \ref{subsec:docintegrals}.

\subsubsection{Additional Declarations}
\label{sec:dec}
Some functions of \Integrals\ require the distinction between small and heavy masses
or loop momenta and external momenta. Thus for the proper function of these functions additional declarations have to be made. Details can be found in section \ref{sub:declarationsintegrals}.

\subsubsection{Representatin of Propagators}
The propagator structure of one-loop integrals like
\be
\label{prop_oneloop}
 \f {1} {(q_1^2-m_1^2)^{n_1}}
\ee
is written in the programme as
\be
\mathtt{AD[\underbrace{\mathtt{den[q_1,m_1],\ldots, den[q_1,m_1]}}_{\rm{n_1\;times}}]}.
\ee
In analogy, the propagator structure of two-loop integrals 
\be
\label{prop_twoloop}
\f {1} {(q_1^2-m_1^2)^{n_1} (q_2^2 -m_2^2)^{n_2} ((q_1+q_2)^2-m_3^2)^{n_3}}
\ee
is written as
\begin{align}
\label{eq:AD}
{\scriptstyle{\mathtt{AD[\underbrace{\scriptstyle\mathtt{den[q_1,m_1],\ldots, den[q_1,m_1]}}_{\rm{n_1\;times}}, \underbrace{\scriptstyle\mathtt{den[q_2,m_2],\ldots,den[q_2,m_2]}}_{\rm{n_2 \;times}},\underbrace{\scriptstyle\mathtt{den[q_1+q_2,m_3],\ldots,den[q_1+q_{2},m_3]}}_{\rm{n_3\;times}}]}}}.
\end{align}
\subsection{Colour Algebra}
Integrals with outgoing gluons or quarks can lead to a quite complicated colour structures.
The following relations can be derived
in the fundamental representation of $SU(N)$ \cite{Field}:
 \begin{align}
 \label{c1}
 f^{bac} \gen^{c}\gen^b&=\f{1}{2} \;i\; N \gen^a,\\
 \gen^c\gen^d f^{dba}f^{acb}&=
 \gen^c\gen^d N \delta^{dc}\nnl
 &=\frac{N^2-1}{2N} N=\frac{N^2-1}{2},\\
 \label{c2}
 \gen^a \gen^e f^{adc}f^{dek} f^{ckb}&=\f{N}{2}\gen^a \gen^e f^{abe}
 =-i \f{N^2}{4}\gen^b.
  \end{align}
 The function \ty{Color} applies eqs.~(\ref{c1}-\ref{c2}) for the special case $N=3$. Note that structure constants
$f^{abc}$ are represented in the programme as \ty{SUNF[a,b,c]}, whereas products of generators $\gen^a \gen^b$ are represented as \ty{SUNT[a,b]}.
\subsection{From Tensor Integrals to Scalar Integrals}
\Integrals\ was originally designed to facilitate the calculation of Wilson coefficients of mass dimension six operators of effective Hamiltonians of the the rare decays $\bsgamma $ and $\bsll$ in the SM and Two-Higgs-Doublet models. In the corresponding integrals two heavy mass scales arise: the top-mass and the $W$-mass (SM) or the charged Higgs mass (THDM). A typical propagator structure is given by 
	\be
	I= \f
		{1}
	{(q_1^2-m_1^2)^{n_1}(q_1^2-m_2^2)^{n_2} ((q_2+k_1)^2 -m_2^2)^{n_3}((q_1+q_2+k_2)-m_2^2)^{n_4}},
	\ee
	where $q_1$ and $q_2$ are the loop momenta, $k_1$ and $k_1$ the external momenta, $n_j \ge0 $ and $\sum_j n_j=6$. The exact calculation of two-loop graphs with two mass scales is technically very demanding. Thus at the moment exact results for diagrams with more than one
	mass scale do not exist beyond one-loop. Therefore the Heavy Mass Expansion (HME) \cite{Smirnov}, an asymptotic expansion in small momenta and masses, is used.
\subsubsection{Heavy Mass Expansion}

\label{sec:taylor}

 The basic idea of the Heavy Mass Expansion (HME) is to use the hierarchy of mass
	scales and momenta to reduce complicated two-loop calculations to simpler ones. The following
	assumptions are made:
	\begin{enumerate}
	\item All the masses of a given Feynman diagram $\Gamma$
	can be divided into a
	set of large \mbox{$\underline{M}=\{M_{1},M_{2}, \ldots$\}} and small
	$\underline{m}=\{m_{1},m_{2}, \ldots\}$ masses.
	\item All external momenta
	$\underline{k}=\{k_{1},k_{2}, \ldots\}$ are small compared to the scale
	of
	the large masses $\underline M$.
	\end{enumerate}
	The ansatz is that the dimensionally regularised (unrenormalised) Feynman integral 
	$F_{\Gamma}$ 
	associated with the Feynman diagram
	$\Gamma$ can be written as
	\be
	\label{hme}
	F_{\Gamma} \stackrel{\underline{M} \to \infty}{\sim}
	\sum_{\gamma} F_{\Gamma / \gamma} \circ
	{\cal T}_{\underline{k}^{\gamma},\underline{m}^{\gamma}} 
	F_{\gamma}(\underline{k}^{\gamma},\underline{m}^{\gamma},\underline{M}),
	\ee
	where the sum is performed over all subgraphs $\gamma$ of $\Gamma$ which 
	fulfil the following two conditions simultaneously:
		\begin{itemize}
	\item $\gamma$ contains all lines with heavy masses
	($\underline{M}$),
	\item $\gamma$ consists of connected\footnote{A graph is called connected when it can not be separated into two or more distinct pieces without cutting any line.} components that are
	one-particle-irreducible with respect to the lines with small masses
	($\underline{m}$).
	\end{itemize}
		
	The operator ${\cal T}$ performs a Taylor expansion in the variables
	$k_i^2/M_j^2$ and $m_l^2/M_j^2$,
	where $k_i$ belongs to $\underline{k}^{\gamma}$, the set of 
	external momenta with respect to the subgraph $\gamma$. $m_l$ belongs to the set of light masses $\underline{m}^{\gamma}$ of 
	$\gamma$. $M_j$ is the heavy mass of the propagator to which the light mass or the external momenta belong to.

\begin{figure}[t]	
\begin{center}	
\label{fig:hme}
\includegraphics[width=13cm]{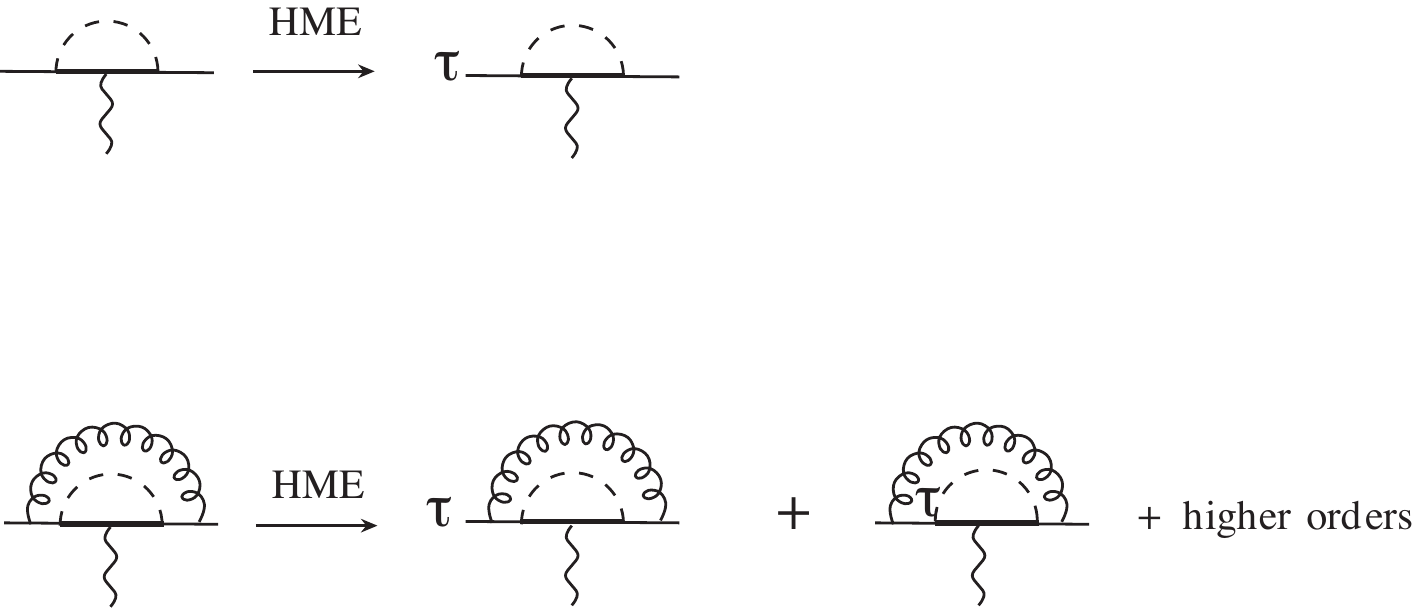}
\end{center}	
\caption{\small
Expansion of the full theory in the HME for the example process $\bsgamma$. $\tau$ symbolises the Taylor expansion in small masses and momenta as described in eq.~(\ref{hme}). Thick lines stand for heavy quarks (in this example the top mass), dashed lines heavy bosons like the $W^\pm, \pi^\pm$ in the SM or the charged Higgs in the THDM. In line two we show the two subdiagrams needed to be evaluated in the HME.} 
\end{figure}	
After Taylor expansion of the two-loop integrals we need to deal with the calculation of a large number of rather simple integrals. 
Matching these results with effective low energy theories we find out to which mass power we must expand the Taylor series in the HME. 
In order to calculate Wilson coefficients of the rare b-decays $\bsgamma$ and $\bsll$ up to \oas-precision we have to match to an effective theory with operators of mass dimension six. Therefore it is sufficient to expand the integrands up to second order in external momenta and small masses. Expansion up to higher order in the external momenta would correspond to Wilson coefficients of operators of higher mass dimensions and can therefore be safely neglected.
The Taylor expansion of the Feynman integrands in external momenta, as well as setting all the light masses to zero, creates
spurious infrared divergences which can be regularised dimensionally. All these divergences cancel out in the matching conditions
relating the full and the effective theory Green functions.

\subsubsection*{Taylor Expansion}
  The expansion in external momenta is performed by the function \\\ty{TaylorExpansion}. It performs the expansion of each propagator in external momenta up to ${\cal O}[($external momenta$)^2/M^2]$
 \begin{align}
 \label{eq:taylorexpansion}
 \f{1}{(q_i+k)^2-M^2}=&
 \f{1}{q_i^2-M^2}
 \left[
 1-\f {k^2 +2kq_i }{q_i^2-M^2}+\f {4 (k q_i)^2 }{(q_i^2-M^2)^2}
 \right] 
 +{\cal O}[k^4/M^4],& \\
 \f{1}{(q_1+q_2+k)^2-M^2}=&
 \f{1}{(q_1+q_2)^2-M^2}&\nonumber \\
 &\left[
 1-\f {k^2 +2k q_1+2 k q_2 }{(q_1+q_2)^2-M^2}+
 \f {4 (k q_1)^2 + 4 (k q_2)^2 +8 q_1 kq_2} {((q_1+q_2)^2-M^2)^2}
 \right]&\nnl
 &+{\cal O}[k^4/M^4],&
 \end{align}
 where
 $q_i\;(i=1,2)$ are the loop momenta, $M$ is a heavy mass and
 $k$ an arbitrary external momentum. \\
 The expansion in small masses up to second order
 \be
 \label{taylormass}
 \f{1}{q_i^2-m^2}=
 \f{1}{q_i^2}
 \left[
 1+\f {m^2 }{q^2}
 \right]+{\cal O}[m^4/q^4],
 \ee
 where $m$ is a small mass, is performed by the function \ty{TaylorMass}. The function expands automatically in all masses not declared as heavy masses with \ty{DeclareHeavyMass}.
\subsubsection*{Scaling}
\label{Scaling}
The routine \ty{Scaling} multiplies all light masses and external momenta with a factor $x$ and sets all terms $x^n$ with $n>2$ to zero.
This is justified in the calculation of Wilson coefficients corresponding to operators of mass dimension six.
Keeping terms with $n>2$ would correspond to the calculation of contributions to Wilson coefficients of higher mass dimensions.

 \subsubsection{Partial Fraction Decomposition and Simplification of Numerators}
 \label{partial}
The routine \ty{PartialFractionOne} (one-loop case) and \ty{PartialFractionTwo} (two-loop case) apply partial
 fraction decompositions. The resulting integrals contain in the propagator denominator a single
 mass parameter together with a
 given loop momentum. 
  \begin{align}
   \label{eq:partial_one}
 \f{1}{(q^2-m_1^2)(q^2-m_2^2)} &= 
 \f{1}{m_1^2-m_2^2} \left[ \f{1}{q^2-m_1^2}-\f{1}{q^2-m^2} \right],\\
 \label{simplification}
 \f{q^2}{(q^2-m_1^2)(q^2-m_2^2)} &= 
 \f{1}{m_1^2-m_2^2} \left[ \f{m_1^2}{q^2-m_1^2}-\f{m_2^2}{q^2-m_2^2} \right]. 
  \end{align}
  Furthermore the routines perform the following relations to get successively rid of loop momenta in the numerator:
 \begin{align}
 \label{red1}
 \f{(q_i^2)^n}{q_i^2-m^2}=&
 (q_i^2)^{n-1} + \f{(q_i^2)^{n-1} m^2}{q_i^2-m^2} \;(i=1,2),&\\
  \label{red2}
 \f {(q_1 q_2)^n} {(q_1^2-m_1^2)(q_2^2-m_2^2)((q_1+q_2)^2-m_3^2)}
 =&
 \f{1}{2} (q_1 q_2)^{n-1}
 \left[
 \f {1}
 {(q_1^2-m_1^2)(q_2^2-m_2^2)}
 \right .\nonumber &\\
 &-
 \f {1} {(q_2^2-m_2^2)((q_1+q_2)^2-m_3^2)} &\nonumber \\
 &-
 \f {1} {(q_1^2-m_2^2)((q_1+q_2)^2-m_3^2)}&
 \nonumber \\
 &+\left .
 \f {m_3^2-m_1^2-m_2^2} {(q_1^2-m_1^2)(q_2^2-m_2^2)((q_1+q_{2})^2-m_3^2)} \right]&
 \end{align}
with $ n\ge 1 $.
In a last step all vanishing massless integrals are set to zero \cite{Muta}:
 \be
 \int d^D q \f{1}{q^2{}^\alpha}=0.
 \ee
 \subsubsection{Tensor Reduction}
 The idea of the tensor reduction is to express tensor integrals in terms of scalar integrals. As integrals over an antisymmetric integrand with symmetric integration boundaries are zero, all integrands with an odd number of loop momenta $q_i ^\alpha, (i={1,2})$ in the nominator can be set to zero before performing the proper tensor reduction.
\label{tensorreduction}
The basic relations for the tensor reduction of one-loop integrals are given by
 \begin{align}
\label{eq:onelooptensor_1}
 \int d^D q \;{q^{\alpha_1} q^{\alpha_2}}{A(q^2)}&=
 \f {1}{D}\int d^D q^2 g^{\alpha_{1} \alpha_{2}}A(q^2),&\\
 \int d^D q \;q^{\alpha_1} q^{\alpha_2} q^{\alpha_3} q^{\alpha_ 4} A(q^2)&=
 \f{1}{D^2 +2 D} &\nnl
 &\int d^D q^4
 (g^{\alpha_1 \alpha_2} g^{\alpha_3 \alpha_4}+
 g^{\alpha_1 \alpha_3} g^{\alpha_2 \alpha_4}+
 g^{\alpha_1 \alpha_4} g^{\alpha_2 \alpha_3})
 A(q^2),&
  \\
 \label{eq:onelooptensor_2}
 \int d^D q\; q^{\alpha_1} q^{\alpha_2} \dots q^{\alpha_{2k}}A(q^2)&=
 \f {\Gamma(2 -\e)}{2 ^k\Gamma(2-\e +k)}
 \int d^D q^ {2k}{ \bf{X^{(k)}}}\; A(q^2),&
  \end{align}
 where
 $A(q^2)$ is an arbitrary scalar function depending on Lorentz invariants of the
 loop momentum $q$ and masses. Usually it is a product of powers of propagators
 \be
 \f {1}{(q^2 -m^2)^{n_1}}
 \ee
times a polynomial of $q^2$.
 $\bf{X^{(k)}}$ stands for permutations of metric tensor components $g^{\alpha_j \alpha_k}$.
 The routine \ty{TensorOne} performs the tensor reduction of one-loop integrals for up to nine Lorentz indices. Results are Taylor expanded in $\e$ up to second order.
 The one-loop relations eqs. ~(\ref{eq:onelooptensor_1}-\ref{eq:onelooptensor_2}) can be generalised to the case of two-loop integrals \cite{Chetyrkin:1998} \cite{Urban:1999tl}
 \begin{flalign}
 \int d^D q_1 d^D q_2 \;\;
 {q_{1}^{\alpha_1} q_{2}^{\alpha_2}}{A(q_1,q_{2})}=&\f {1}{D} \int d^D q_1 d^D q_2 \;A(q_1,q_{2}) \;(q_1\cdot q_2) {g^{\alpha_{1} \alpha_{2}}},\\
 \int d^D q_1 d^D q_2\;\; A(q_1,q_{2})
 q_1^{\alpha_{1}} q_1^{\alpha_{2}} q_1^{\alpha_{3}} q_2^{\alpha_{4}} =&\f {1}{D^2 +2 D}
 \int d^D q_1 d^D q_2\; \;A(q_1,q_{2}) q_1^2 \;(q_1\cdot q_2) \nonumber \\
 &
 (g^{\alpha_1 \alpha_2} g^{\alpha_3 \alpha_4}+
 g^{\alpha_1 \alpha_3} g^{\alpha_2 \alpha_4}+
 g^{\alpha_1 \alpha_4} g^{\alpha_2 \alpha_3})
 \\
 \int d^D q_1 d^D q_2\;A(q_1,q_{2})
 {q_1^\alpha q_1^\beta q_2^\gamma q_2^\delta
 }
 =&
 \f {1}{D^3 +D^2- 2D}
 \int d^D q_1 d^D q_1\; A(q_1,q_{2})\nonumber \\
 &\bigl[
 \left (
 (1+D)
 {q_1^2 q_2^2}
 -2 {(q_1 \cdot q_{2})^2}
 \right) 
 g^{\alpha_1 \alpha_2} g^{\alpha_3 \alpha_4}
  \nnl
 &+\left(
 -{q_1^2 q_2^2+ D} {(q_1 \cdot q_{2})^2}
 \right)
  g^{\alpha_1 \alpha_3} g^{\alpha_2 \alpha_4} 
  \nnl
  &+
 g^{\alpha_1 \alpha_4} g^{\alpha_2 \alpha_3}) 
 \bigr ], 
 \end{flalign}
 where $A(q_1,q_{2})$ is an arbitrary scalar function of $q_1$ and $q_2$ and arbitrary masses.
 It is usually a product of powers of propagators
 \be
\label{example}
 \f {1}{(q_1^2-m_1^2)^{n_{1}}(q_2^2-m_2^2)^{n_{2}}((q_1+q_2)^2-m_3^2)^{n_{3}}}
 \ee
 times a polynomial in $q_1^2$, $q_1^2$, $q_1 q_2$, but the concrete form of this function has no importance for the tensor reduction.
 The function \ty{TensorTwo} performs the two-dimensional tensor reduction for up to four Lorentz indices.
In the case of factorising integrals corresponding to $c=0$ in the integrand of eq.~(\ref{example}), \ty{TensorTwo} performs a one-dimensional tensor reduction calling \ty {TensorOne}. From the tensor reduction we obtain additional terms of $q_1^2, \;q_2^2 $ or $ q_1 q_2 $ in the numerator. This makes a
subsequent usage of the identities \ty{PartialFractionOne} and \ty{PartialFractionTwo}, described in section \ref{partial}, necessary. 

\subsubsection{Substitutions}
 The function \ty{Substitutions} makes substitution in the integrands of factorising two-loop-integrals such that the propagator structure contains no overlapping loop momenta by applying the following relations
\begin{align}
 \int d^D q_1 d^D q_2 
 { \textstyle{
 \dfrac {S(q_{1},q_{2})} {(q_1^2-m_1^{2})^{n_1} ((q_1+q_2)^2-m_2^2)^{n_3}}
 }}
  & =
  \int d^D q_1 d^D q_2 \dfrac {S(q_{1},q_{2}-q_{1})} {(q_1^2-m_1^{2})^{n_1} (q_2^2-m_2^2)^{n_3}},&\\
  \int d^D q_1 d^D q_2 \f {S(q_{1},q_{2})} {(q_2^2-m_1^{2})^{n_2} ((q_1+q_2)^2-m_2^2)^{n_3}}
  & =
  \int d^D q_1 d^D q_2 \f {S(q_{1}- q_{2},q_{2})} {(q_2^2-m_1^{2})^{n_1} (q_1^2-m_2^2)^{n_3}},&
  \label{eq:substitution1}
 \end{align}
where $S(q_{1},q_{2})$ is a polynomial in $q_{1}^{2}$, $q_{2}^{2}$ and $q_{1} \cdot q_{2}$.
A subsequent final partial fraction of the so obtained integrands leads then to the desired scalar integrals.

\subsubsection{Transforming the Propagator Structure}
\label{substitution}
The routine \ty{SimplifyPropagator} transforms the propagator structure of a scalar loop integrals to the forms needed for the loop integration functions.
Thus the propagator structure of non-factorising scalar two-loop integrals eq.~({\ref{eq:AD}})
is transformed to the short form
\be
\mathtt {G[i[m_1,n_1],i[m_2,n_2],i[m_3,n_3]]}.
\ee
The routine replaces the propagator structure of factorising two-lop integrals
\be
 \mathtt{AD[\underbrace{\mathtt{den[q_1,m_1],...., den[q_1,m_1]}}_{\rm{n_1\;times}}, \underbrace{\mathtt{den[q_2,m_2],....,den[q_2,m_2]}}_{\rm{n_2 \;times}}]}
\ee
with
\be
\mathtt{AD[i[m_1,n_1],i[m_2,n_2]]}
\ee
and the propagator structure of one-loop integrals
\be
 \f {1} {(q_1^2-m_1^2)^{n_{1}}}
\ee
by
\be
{\mathtt{AD[i[m_1,n_1]]}}.
\ee
\ty{SimplifyPropagator} orders furthermore scalar two-loop integrals with one vanishing mass in the denominator in such a way that the propagator denominator with overlapping loop momenta has always no additional mass term ($m_3 =0$):
\begin{align}
 \int d^D q_1 d^D q_2 \f {1} {(q_1^2)^{n_3} (q_2^2-m_2^2)^{n_2} ((q_1+q_2)^2-m_1^2)^{n_1}}
 \nonumber \\
 =
\int d^D q_1 d^D q_2
 \f {1} {(q_1^2-m_2^2)^{n_2} (q_2^2)^{n_3} ((q_1+q_2)^2-m_1^2)^{n_1}} \nonumber \\
=
\int d^D q_1 d^D q_2
\f{1} {(q_1^2-m_1^2)^{n_1} (q_2^2-m_2^2)^{n_2} ((q_1+q_2)^2)^{n_3}}.
\label{finalform}
 \end{align}
 The last line of eq.~(\ref{finalform}) is the ordering of propagator denominators needed for the following loop integration routines. This ordering is necessary, as the two-loop integrals are represented in the programme as a non-commuting list.
 
\subsection{Loop Integration of Scalar One Loop Integrals}
\label{sec:Loop_Integration_One}
After Taylor expansion, tensor reduction and subsequent partial fractions the one-loop tensor integrals are transformed to a bigger number of scalar integrals proportional to \cite{Bobeth:1999ww}:
 \begin{align}
 \mu^{2 \e}\int \f{d^{D}\;q}{(2 \pi)^{- 2 \e}}\f{1}{(q^2-m^2)^n}
  &=
  \f{\mu^{2 \e}} {(2\pi) ^{- 2 \e}} \f {\pi^{D/2} \; \Gamma(1+\e) } { (m^2)^{n-D/2}} \;
 C^{(1)}_n
  \nonumber \\
  &=
 \f {\pi^2}{(m^2)^ {n-2}} \;
 \left(\left(\f{\mu^{2}} {m^2}\right)^\eps \;2 ^{2 \eps}\pi^{\eps} \; \Gamma(1+\e)\right) \;
 C^{(1)}_n
 \nonumber \\
 &=
  \f{\pi^2}{ (m^2)^ {n-2}} N_{\e}^{(1)}(m)\; C^{(1)}_n %
 \label{intoneloop},
 \end{align}
 where for arbitrary $n$ and $m$ \cite{Collins}:
 \begin{align}
\label{Ne}
 N_{\e}^{(1)}(m)&=
 \left(\f{\mu^{2}} {m^2}\right) ^{ \e} 2 ^{2\e}\pi^{\e} \; \Gamma(1+\e)
 \nonumber \\
 &=
 1-\e \kappa + \e^2 \left(\f{1}{12}\pi^2 +\f{1}{2}\kappa(m)^2\right)+{\cal{O}}(\e)^3, \\
 \kappa(m) &=\gamma_E -\ln (4 \pi) +\ln \f{m^2}{\mu^2}, \\
 \label{oneloop}
 C^{(1)}_n &= i \f{(-1)^n}{(n-1)!} (1+\e)_{n-3},
	\end{align}
	which vanishes for $n \le 0$.
 In eq.~(\ref{oneloop}) we introduced the Pochhammer symbol
 \be
 (a)_k = \f{\Gamma(a+k)}{\Gamma(a)}= \left\{ \begin{array}{cc}
 a(a+1)(a+2)...(a+k-1), & k \geq 1,\\
 1,   & k = 0,\\
 1/[(a-1)(a-2)...(a-|k|)], & k \leq -1
 \end{array} \right.
 \ee
 for integer $k$ and complex $a$. The prefactors of $C^{(1)}_n$ are chosen such that
 $C^{(1)}_n$ is free of common factors of the one-loop integration. The factor $N_{\e}^{(1)}(m_1)$ summarises the $\e$-dependent part of the common prefactors.
 The function \ty {ScalIntOne} performs the scalar one-loop integration by replacing the propagator structure \ty{AD[i[m,n]]} by the right hand side of eq.~(\ref{intoneloop}):
 \be
 \mathtt{AD[i[m,n]]} \rightarrow\ \f{\pi^2}{ (m^2)^ {n-2}} \mathtt{Ne[m]\; C^{(1)}_n},
 \ee
where \ty{Ne[m]} corresponds to eq.~(\ref{Ne}) and $\mathtt{C^{(1)}_n}$ to eq.~(\ref{oneloop}) up to second order in \ty{eps}. The final result is expanded in \ty{eps} up to first order. 
 
\subsection{Loop Integration of Scalar Two Loop Integrals}	
\subsubsection{Recurrence Relations}
\label{sec:recurrence}
In this section we will show how to reduce scalar two loop integrals independent of external momenta to master integrals, where the highest power of all appearing propagators is one. These master integrals can then be automatically integrated.

  We will first give a general derivation of the recurrence relations for arbitrary masses $ m_1, \;m_2$ and $m_3$ for the integral
  \begin{align}
\label{general}
  G_{n_1,n_2,n_3}^{m_1,m_2,m_3}
 &\equiv
  \f {\mu^{4 \e}}{(2 \pi)^{-4 \e}}\int d^D q_1 d^D q_2
 \f {1} {(q_1^2-m_1^2)^{n_1} (q_2^2 -m_2^2)^{n_2} ((q_1+q_2)^2-m_3^2)^{n_3}}.
 \end{align}
Its propagator structure
is written in the package as $\mathtt {G[i[m_1,n_1],i[m_2,n_2],i[m_3,n_3]]}$.
 The derivation of the recurrence relations starts with the following identities \cite{DT93}:
 \begin{align}
 \label{rec00}
 \int d^D q_1 d^D q_2 \f{\partial}{\partial q_1 ^\mu}
 \left(
 \f
 {q_1^{\mu}}
 {(q_1^2-m_1^2)^{n_1} (q_2^2 -m_2^2)^{n_2} ((q_1+q_2)^2-m_3^2)^{n_3}}
 \right)
 &=0,\\
 \label{rec11}
 \int d^D q_1 d^D q_2 \f{\partial}{\partial q_2 ^\mu}
 \left(
 \f
 {q_2^{\mu}}
 {(q_1^2-m_1^2)^{n_1} (q_2^2 -m_2^2)^{n_2} ((q_1+q_2)^2)-m_3^2)^{n_3}}\right)
 &=0,\\
 \label{rec22}
 \int d^D q_1 d^D q_2 \f{\partial}{\partial{q_1 ^\mu}}
 \left(
 \f
 {q_1^{\mu}}
 {(q_1^2-m_1^2)^{n_1} ((q_1+q_2)^2 -m_2^2)^{n_2} (q_2^2-m_3^2)^{n_3}}
 \right)
 &=0.
 \end{align}
 Substitutions of the integration variables in
 eq.~(\ref{rec00}) lead to eqs.~((\ref{rec11})-(\ref{rec22})).
 With the help of the the Gaussian integral theorem we can transform the integral to a vanishing surface integral with symmetric boundaries and an asymmetric integrand.
 In oder to simplify the notation we will
 use
 \be
  G_{n_1,n_2,n_3}^{m_1,m_2,m_3}\equiv G_{n_1,n_2,n_3}
 \ee
 in this section.
 From eq.~(\ref{rec00}) we get
 \begin{align}
 \label{finals1}
 (D- 2 n_1 -n_3)G_{n_1,n_2,n_3}=& 2 n_1 m_1^2 G_{n_1+1,n_2,n_3}+
 n_3 (G_{n_1-1,n_2,n_3+1}-G_{n_1,n_2-1,n_3+1})&\nonumber \\
 &+n_3 (m_1^2 -m_2^2+m_3^2)G_{n_1,n_2,n_3+1}.&
 \end{align}

 From eq.~(\ref{rec11}) or directly by replacing $ n_1\leftrightarrow n_2$ and $ m_1\leftrightarrow m_2$
 in eq.~(\ref{finals1}) we get
 \begin{align}
 \label{finals2}
 (D- 2 n_2 -n_3)G_{n_1,n_2,n_3}=& 2 n_2 m_2^2 G_{n_1,n_2+1,n_3}+
 n_3 (G_{n_1,n_2-1,n_3+1}-G_{n_1-1,n_2,n_3+1})&\nonumber \\
 &+n_3 (m_2^2 -m_1^2+m_3^2)G_{n_1,n_2,n_3+1}.&
 \end{align}
 From eq.~(\ref{rec22}) we obtain
 \begin{align}
 \label{finals3}
 (D- 2 n_1 -n_2)G_{n_1,n_2,n_3}=& 2 n_1 m_1^2 G_{n_1+1,n_2,n_3}+
 n_2 (G_{n_1-1,n_2+1,n_3}-G_{n_1,n_2+1,n_3-1})&\nonumber \\
 &+n_2 (m_1^2 +m_2^2-m_3^2)G_{n_1,n_2+1,n_3}.&
 \end{align}
 The last three equations connect integrals with the sum of powers $n_1+n_2 +n_3$ with integrals
 where the sum of the powers is lowered by 1. They form an equation system, which can be used to extract the integrals\\ $ G_{n_1+1, n_2,n_3},\; G_{n_1, n_2+1,n_3}\;G_{n_1, n_2,n_3+1}$.
 Solving it we obtain the following recurrence relations \cite{DT93}: 
  \begin{align}
 \label{pippo1}
 G_{n_1+1,n_2,n_3}=& \f{1}{n_1 \;m_1^2\;\Delta (m_1,m_2,m_3)}\nonumber \\
 &
 \bigl \{
 \bigl[
 n_2 \;(m_1^2 -m_3^2 )(m_1^2 -m_2^2 +m_3^2)+n_3 \;(m_1^2 -m_2^2)(m_1^2+m_2^2-m_3^2)
 \nnl
 &
 \left. 
 + D \;m_1^2 \;(-m_1^2 +m_2^2 +m_3^2)-n_1\; \Delta\;(m_1,m_2,m_3) 
 \right]
 G_{n_1,n_2,n_3}
 \nnl
 &+n_2 \;m_2^3 (m_1^2 -m_2^2 +m_3^2)
 \left[
 G_{n_1,n_2+1,n_3-1}-G_{n_1-1,n_2+1,n_3}
 \right]
 \nnl
 &+
 n_3 \;m_3^2 (m_1^2 +m_2^2 -m_3^2)
 [
 G_{n_1,n_2-1,n_3+1}-G_{n_1-1,n_2,n_3+1}
 ]
 \bigr \}
 \end{align}
 with the determinant of the corresponding equation system
 \be
 \Delta (m_1,m_2,m_3)=2(m_1^2m_2^2 +m_1^2 m_3^2 +m_2^2 m_3^2)-(m_1^4+m_2^4+m_3^4).
 \ee

 Replacing $n_1\leftrightarrow n_2$ and $m_1\leftrightarrow m_2$ in
  eq.~(\ref{pippo1}) we get
 \begin{align}
 \label{heidi2}
 G_{n_1,n_2+1,n_3}=& \f{1}{n_2\; m_2^2\;\Delta (m_1,m_2,m_3)}\nonumber \\
 &
 \bigl \{
 \left[
 n_1 \;(m_2^2 -m_3^2 )(m_2^2 -m_1^2 +m_3^2)+n_3 \;(m_2^2 -m_1^2)(m_1^2+m_2^2-m_3^2)
 \right.\nonumber \\
 &
 \left. 
 +D \;m_2^2 \;(-m_2^2 +m_1^2 +m_3^2)-n_2 \;\Delta(m_1,m_2,m_3) \right]G_{n_1,n_2,n_3}\nonumber \\
 &+n_1\; m_1^2 (m_2^2 -m_1^2 +m_3^2)
 \left[
 G_{n_1+1,n_2,n_3-1}-G_{n_1+1,n_2-1,n_3}
 \right]\nonumber \\
 &+
 n_3 \;m_3^2 (m_1^2 +m_2^2 -m_3^2)
 \left[
 G_{n_1-1,n_2,n_3+1}-G_{n_1,n_2-1,n_3+1}
 \right] \bigr \} .
 \end{align}
 In analogy, we get by replacing $n_1\leftrightarrow n_3$ and $m_1\leftrightarrow m_3$ in eq.~(\ref{pippo1})
  \begin{align}
 \label{heidi3}
 G_{n_1,n_2,n_3+1}=& \f{1}{n_3 \;m_3^2\;\Delta (m_1,m_2,m_3)}\nonumber \\
 &
 \bigl\{
 \left[
 n_1 (m_3^2 -m_2)^2 )(m_2^2 -m_1^2 +m_3^2)+n_2 (m_3^2 -m_1^2)(m_1^2+m_3^2-m_2^2)
 \right.\nonumber \\
 &
 \left. 
 +D\; m_3^2 (-m_3^2 +m_1^2 +m_2^2)-n_3\; \Delta(m_1,m_2,m_3) \right]G_{n_1,n_2,n_3}\nonumber \\
 &+n_1 \;m_1^2 (m_2^2 -m_1^2 +m_3^2)
 \left[
 G_{n_1+1,n_2-1,n_3}-G_{n_1+1,n_2,n_3-1}
 \right]\nonumber \\
 &+
 n_2 \;m_2^2 (m_1^2 +m_3^2 -m_2^2)
 \left[
 G_{n_1-1,n_2+1,n_3}-G_{n_1,n_2 +1,n_3-1}
 \right] \bigr\}.
 \end{align}
 The general recurrence relations eqs~(\ref{pippo1}), (\ref{heidi2}) and (\ref{heidi3}) are
 implemented in the rule \ty{recurrence} which is part of the integration routine \ty{ScalIntTwoThreeMasses}.
 \subsubsection*{Recurrence Relations for Scalar Integrals with One Massless Propagator}
 \label{recurrenceb}
In the following we consider the special case, that one of the masses in eq.~(\ref {pippo1})
vanishes.
 Without restrictions we can choose this mass to be $m_3$. Taking the limit $m_3\rightarrow 0$ we get from eqs.~(\ref {pippo1}) and (\ref {heidi2})
 \be
 \begin{array}{ccrlrl}
 G_{n_1+1,n_2,n_3}^{m_1,m_2,0} &=& \f{1}{m_1^2 \;n_1 (1-x)}
 & \multicolumn{3}{l}{
 \left\{ [ D-n_1-n_2-n_3 + x (n_1 -n_3)] G_{n_1 n_2 n_3}^{m_1,m_2,0}
 \right.} \vspace{0.2cm} \\ &&&
 + & x n_2 & \left. \left[ G_{n_1-1,n_2+1,n_3}^{m_1,m_2,0}
    -G_{n_1,n_2+1,n_3-1}^{m_1,m_2,0} \right] \right\},
 \end{array}
 \label{rec11b}
 \ee
 \be
 \begin{array}{ccrlrl}
 G_{n_1,n_2+1,n_3}^{m_1,m_2,0} &=& -\f{1}{m_2^2 n_2 x (1-x)}
 & \multicolumn{3}{l}{
 \left\{ [ x (D-n_1-n_2-n_3) + n_2 -n_3] G_{n_1 n_2 n_3}^{m_1,m_2,0}
 \right.} \vspace{0.2cm} \\ &&&
 + & n_1 & \left. \left[ G_{n_1+1,n_2-1,n_3}^{m_1,m_2,0} \right]\right\},
 \end{array}
 \label{rec22b}
 \ee
 where $x = m_2^2/m_1^2$ \cite{DT93}.
 From eq.~(\ref{heidi3}) we see that the limit $m_3 \rightarrow 0$ does not exist for $G_{n_1,n_2,n_3+1}^0$.
 The recurrence relation for $G_{n_1,n_2,n_3+1}^0$ in this limit can be derived from eq.~(\ref{pippo1}) by eliminating $G_{n_1+1,n_2,n_3}^0$ with the help of eq.~(\ref{heidi2}) and $G_{n_1,n_2+1,n_3}^0$ with the help of eq.~(\ref{heidi3}). Thus we obtain
 \be
 \begin{array}{ccrlrl}
 G_{n_1, n_2, n_3+1}^{m_1,m_2,0} &=& \f{1}{m_1^2 n_3 (1-x)^2}
 & \multicolumn{3}{l}{
 \left\{ [(1+x)(-D) + 2 n_2 + (1+3x) n_3] G_{n_1 n_2 n_3}^{m_1,m_2,0}
 \right.} \vspace{0.2cm} \\ &&&
 + & 2 x n_2 & \left[ G_{n_1,n_2+1,n_3-1}^{m_1,m_2,0}
    -G_{(n_1-1)(n_2+1)n_3}^{m_1,m_2,0} \right]
 \vspace{0.2cm} \\ &&&
 + & (1-x) n_3 & \left. \left[ G_{n_1(n_2-1)(n_3+1)}^{m_1,m_2,0}
    -G_{(n_1-1)n_2(n_3+1)}^{m_1,m_2,0} \right] \right\}.
 \end{array}
 \label{rec33}
 \ee
 Eqs.~(\ref{rec11b}-\ref{rec33}) are implemented in the rule \ty{recurrenceb}. This rule is part of the integration routine \ty{ScalIntTwo}.

\subsubsection{Loop Integration of Master Integrals}
\label{sec:Loop_Integration_Two}
In the last section we have shown how to reduce scalar two-loop integrals independent of external momenta to master integrals. This section will focus on the loop integration of special cases of these master integrals. We will focus on integrals with only two different masses ($m_1=m_3$) and the case of one vanishing mass ($m_3=0$) in eq.~(\ref{general}).

\subsubsection*{Scalar Two loop Integrals with Two Different Masses}
\label{stauintegrals}
The routine \ty{ScalIntTwoThreeMasses} can automatically perform the integration of scalar two-loop integrals of the type of eq.~(\ref{general})
for the special case $m_1=m_3$.
In a first step the integrands are ordered by making the following substitutions
\begin{align}
 & \int \tfrac{d^D q_1 d^D q_2} {(2 \pi)^{-4 \e}}
 \tfrac {1} {(q_1^2-m_1^2)^{n_1} (q_2^2-m_1^2)^{n_2} ((q_1-q_2)^2-m_2^2)^{n_3}}\nnl
=&\int \tfrac{d^D q_1 d^D q_2} {(2 \pi)^{-4 \e}}
\tfrac {1} {(q_1^2-m_2^2)^{n_3} (q_2^2-m_1^2)^{n_1} ((q_1+q_2)^2-m_1^2)^{n_2} }\nnl
=&
\label{subaxino}
\int \tfrac{d^D q_1 d^D q_2} {(2 \pi)^{- 4 \e}}
 \tfrac {1} {(q_1^2-m_1^2)^{n_1} (q_2^2-m_2^2)^{n_3} ((q_1+q_2)^2-m_1^2)^{n_2}},
\end{align}
where the order of the integrands given in the last line of eq. (\ref{subaxino}) is the order needed for the following loop integration.

With the help of the recurrence relations eqs.~(\ref{pippo1}), (\ref{heidi2}) and (\ref{heidi3}) we can reduce integrals of the form eq. (\ref{subaxino}) to the following master integral:
  \begin{align}
  \label{masteri} G_{1\;\;\;\;1\;\;\;\;1}^{m_1\; m_2\;m_1}
&=
  \f {\mu^{4 \e}}{(2 \pi)^{-4 \e}}\int d^D q_1 d^D q_2
 \f {1} {(q_1^2-m_1^2)^{n_1} (q_2^2 -m_2^2)^{n_2} ((q_1+q_2)^2-m_1^2)^{n_3}}&\nonumber \\
 &=
 \pi^4 \;m_1^2\,N_{\e}^{(2)}(m_1)\,C_{1\;\;\;1\;\;\;1}^{m_1m_2m_1\;(2)},&
  \end{align}
where $N_{\e}^{(2)}(m_1)$ collects all $\e$-dependent parts of the common prefactors of the two-loop integrals.
It is given by
\begin{align}
 N_{\e}^{(2)}(m_1)&=(N_{\e}^{(1)}(m_1))^2=
 \left(\f{\mu^{2}} {m_1 ^{2}}\right)^{2\e} 2 ^{4 \e}\pi^{2\e} \; \Gamma(1+\e)^2&
 \nonumber \\
 &=
	 \label{ne2}
 1- 2 \e \kappa(m_1) + \e^2 \left(\f{1}{6}\pi^2 +2 \kappa(m_1)^2\right)+{\cal{O}}(\e)^3.&
 \end{align}
 
and $C_{1\;\;\;1\;\;\;1}^{m_1m_2m_1\;(2)}$ by \cite{DT93}
\begin{align}
\label{eq:masteraxino}
C_{1\;\;\;1\;\;\;1}^{m_1m_2m_1\;(2)}&=
\f{1}{(1-\e) (1-2\e)}
\left[
-\frac{1}{\e^2} \left({1 + \frac{x}{2}}\right) 
\right. &
\nnl
&\left . +
 \frac {1}{\e}
\left({x\,\log(x)}\right)-
\frac{1}{2}
 \left({x \,{\log(x)}^2}\right) +
 \left( 2 - \frac{x}{2} \right) \,\phi(x)
\right].&
\end{align}
The function $\phi(x)$ depends on the mass relation between $m_{1}$ and $m_{2}$. 
\begin{itemize}
\item If
\be
\label{eq:x}
0<x=\f{m_2^2}{m_1^2} < 1,
\ee
then $\phi(x)$ is given by
\be
\phi(x)= 4 \sqrt{\f{x}{4-x}}\;\cl_2 \;\left(2 \arcsin\left(\f{\sqrt{x}}{2} \right)\right),
\ee
where $\cl_2$ is Clausen's integral function \cite{Abro}
\begin{align}
\cl_2(\theta)
&=
S_2(\theta)= \Im[\li_2(e^{i \theta})]=
-\int_0^{\theta} dt \ln\left| 2 \sin \left(\f{t}{2}\right)\right |
\nnl
.
\end{align}

\item If $x>1$ then 
\be
\phi(x)= 
\f{1}{\lambda(x)}
\left[
-4 \li_{2}
\left(
\f{1-\lambda(x)}{2}
\right)
 +2 \ln^{2} \left(\f{1-\lambda(x)}{2}\right)
 -\ln^{2}(x)+\f{\pi^{2}}{3}
\right],
\ee
\end{itemize}
where 
\be
\lambda(x)=\sqrt{1-\f{4}{x}}.
\ee

\subsubsection*{Scalar Two Loop Integrals with One Mass Scale}
If in eq.~(\ref{masteri}) all masses are equal we obtain \cite{DT93}
  \be
  G_{1\;\;\;\;1\;\;\;\;1}^{m_1\; m_1\;m_1}
 =
 \pi^4 \;m_1^2\,N_{\e}^{(2)}(m_1)\,C_{1\;\;\;1\;\;\;1}^{m_1m_1m_1\;(2)},
  \ee
where
\be
\label{eq:simpleclausen}
 C_{1\;\;\;1\;\;\;1}^{m_1m_1m_1\;(2)}=\f{1}{(1-\e)(1-2 \e)} 
 \left(
 -\f{3}{2 \e^{2}} +2 \sqrt{3} \cl_{2} \left(\f{\pi}{3}\right),
 \right).
 \ee
 where $\cl_{2} \left({\pi}/{3}\right)=1.0149417...$ is the maximum of Clausen's integral\cite{DT93}.
The function \ty{ScalIntTwoThreeMasses} applies the substitutions of eq.~(\ref{subaxino}) and the recurrence relations eqs.~(\ref{pippo1}), (\ref{heidi2}) and (\ref{heidi3}) to propagator structures of the form\\
$\mathtt{G[i[m_1,n_1],i[m_2,n_2],i[m_1,n_3]]}$. This leads to numerous terms proportional to\\ $\mathtt{G[i[m_1,1],i[m_2,1],i[m_1,1]]}$, which can be replaced by the master integral (\ref{eq:masteraxino}):
\begin{align}
\mathtt{G[i[m_1,1],i[m_2,1],i[m_1,1]]}
 \rightarrow
 \pi^4 \;m_1^2\,\mathtt{N2[m1])}\,
 C_{111}^{(2)},
\end{align}
where $\mathtt{N2[m1]}$ correspond to eq.~(\ref{ne2}) up to second order in \ty{eps} and $ C_{111}^{(2)}$ to eq.~(\ref{eq:masteraxino}) for $m_{1} \neq m_{2}$ and to 
 eq.~(\ref{eq:simpleclausen}) for $m_{1}=m_{2}$.
 The final result is expanded up to zeroth order in \ty{eps}.
\subsubsection*{Scalar Two Loop Integrals with One Massless Propagator}
The function \ty{ScalIntTwo} is able to perform the loop integration for integrals of type eq.~(\ref{general}), if one of the three masses in the propagators is zero. We can choose this to be $m_3$, as all other cases can be tranformed to this special case with the help of eq.~(\ref{finalform}) by the routine \ty{SimplifyPropagator}. Then we get for the D-dimensional two-loop integral \cite{Bobeth:1999ww}
 	 \begin{align}
 	G_{n1,n2,n3}^{m_1,m_2,0}
	&= \f{\mu ^{4 \e}}{(2 \pi)^{-4 \e}}
 	\int \f{d^{D}q_1 \; d^{D}q_2}{(q_1^2-m_1^2)^{n_1}
  	  (q_2^2-m_2^2)^{n_2}[(q_1-q_2)^2]^{n_3}}
  	  \nnl
 &=	
 \f{\mu ^{4 \e}\pi^{D}}{(2 \pi)^{-4 \e}} 
 \f{ \Gamma(1+\e)^2 }{(m_1^2)^{n_1+n_2+n_3-D}}
 C^{(2)}_{n_1 n_2 n_3}
 \nnl	
 &=	
 \f{\pi	^4}{(m_1^2)^{n_1+n_2+n_3-4} }
 \left(\left(\f{\mu^2}{m_1^2}\right)^{2\e} 2 ^{4 \e}\pi^{2 \e} \; \Gamma(1+\e)^2\right) \;
  C^{(2)}_{n_1 n_2 n_3}
 \nnl
 &=
 \f{\pi^4}{(m_1^2)^{n_1+n_2+n_3-4} } N_{\e}^{(2)}(m_1)\; C^{(2)}_{n_1 n_2 n_3}
 \label{int1}
 \end{align}
 with arbitrary integer powers $n_1$, $n_2$ and $n_3$ and with
 $m_1$ and $m_2 \neq 0$. All the two-loop integrals defined in eq.~(\ref{int1}) vanish when either $n_1$ or $n_2$ is non-positive.

 Performing the integration in eq.~(\ref{int1}), we have to distinguish the following cases of non-vanishing integrals:
\begin{enumerate}
 \renewcommand{\labelenumi}{\alph{enumi})}
\item two of the masses are equal,
\item the second mass $m_2$ vanishes,
\item the masses $m_1$ and $m_2$ are different,
\item one of the powers $n_i \;(i={1,2,3})$ is zero (factorising two-loop integrals).
\end{enumerate}
As the first three cases have the prefactor $N_{\e}^{(2)}(m_1)/((m_1^2)^{n_1+n_2+n_3-4}) $ in common, we will only display the corresponding values of $ C^{(2)}_{n_1 n_2 n_3}$:

\begin{enumerate}
 \renewcommand{\labelenumi}{\alph{enumi})}
 \item With the help of Feynman-parameterisation \cite{Bobeth:1999ww} we get for two equal masses $m_1=m_2$ from eq. (\ref{int1})
 \be
 \label{page55}
 C^{(2)}_{n_1 n_2 n_3} =
 (-1)^{n_1+n_2+n_3+1} \f{(2-\e)_{-n_3} (1+\e)_{n_1+n_3-3}(1+\e)_{n_2+n_3-3}
 }{(n_1-1)! (n_2-1)! (n_1+n_2+n_3-4+2\e)_{n_3}}.
 \ee
 \item
 	If in eq.~(\ref{int1}) the second mass $m_2$ vanishes, we again derive with the help of Feynman-parameterisation
 \begin{align}
 \label{page68}
 C^{(2)}_{n_1 n_2 n_3} &=
 (-1)^{n_1+n_2+n_3+1}\nnl
  &\dfrac{ 
  (1+2\e)_{n_1+n_2+n_3-5} (1+\e)_{n_2+n_3-3} (1-\e)_{1-n_2} (1-\e)_{1-n_3}
  }
  { (n_1-1)! (n_2-1)! (n_3-1)! (1-\e)(1-\f{1}{3}\pi^2\e^2 + {\cal O}(\e^3))}.
 \end{align}
\item
If $m_{1} \neq m_2$ and none of the two
 masses vanishes, the routine
 \ty {ScalInt} reduces
all integrals with three positive indices to a term proportional to the master integral $G_{1\;\;\;\;1\;\;\;\;1}^{m_1\; m_2\;0}$
 with the help of recurrence relations eqs.~(\ref{rec11b}-\ref{rec33}). The corresponding $C^{(2)}_{111}$ is given by
 \begin{align}
 \label{masterintegral}
 C^{(2)}_{111} =&
 \f{1}{2 (1-\e) (1-2\e)} \nnl
 &
 \left[-\f{1+x}{\e^2} \;+\; \f{2}{\e} x \ln x \;+\;
 (1-2x) \ln^2 x \;+\; 2 (1-x) \rm{Li}_2\left(1-\f{1}{x}\right) \;+\; {\cal O}(\e)\right],
 \end{align}
where the dilogarithm $\li_2$ is given by \cite{Field}
\begin{align}
\label{dilog}
\li_2(x)&= -\int_0^x \f {\ln (1-t)} {t}dt = -\int_0^1 \f {\ln (1-x t)} {t}dt\nonumber \\
 &= -\int_{1-x}^1 \f {\ln (t)} {1-t}dt=\int_0^1 \f {\ln (t)} {t-1/x}dt
\end{align}
and $x$ by eq.~(\ref{eq:x})\footnote{
The definitions of eq.~(\ref{dilog}) correspond to the definitions used in \Mathematica~\cite{Mathematica}.
Unfortunately this is not the case for the conventions used in \Maple~\cite{Maple}.
Both conventions are related by
\be
\li_2 ^{\text{\Mathematica}} (1-x)= \li_2 ^{\text{\Maple}}(x)\nonumber.
\ee
}
\item When two indices are positive, but one of the $n_i$ in eq.~(\ref{int1}) equals zero, the two-loop integrals reduce to products of one-loop integrals. Without restriction we can choose $ n_3=0$ and obtain from eq.~(\ref{intoneloop})
\begin{align}
\label{fact}
 G_{n1,n2,0}^{m_1,m_2,0}
 &=
  \f{\pi^4 }{(m_1^2)^ {n_1+n_2-4}} \;N_{\e}^{(1)}(m_1)N_{\e}^{(1)}(m_2)\;C^{(1)}_{n_1} C^{(1)}_{n_2},
\end{align}
where
\be\label{facintc}
C^{(1)}_{n_1} C^{(1)}_{n_2}
=-\f{(-1)^{n_1} (-1)^{n_2}}{(n_1-1)!(n_2-1)!} (1+\e)_{n_1-3}(1+\e)_{n_2-3}.
\ee
If $m_1 \neq m_2$ we get
\be
\label{ne11}
N_{\e}^{(1)}(m_1)N_{\e}^{(1)}(m_2)={N_{\e}^2(m_1)}
  \;\left(1-\ln(x) \;\e +\f{1}{2}\;\ln^2(x)\; \e^2 \right) +\ope(\e^3)
\ee
with $x$ defined in eq.~(\ref{eq:x}). If both masses equal $m_1$, we simply have
\bea
(N_{\e}^{(1)}(m_1))^2=N_{\e}^{(2)}(m_1)
\eea
defined in eq.~(\ref{ne2});
if both masses equal $m_2$ we derive
\be
\label{ne22}
 (N_{\e}^{(1)}(m_2))^{2} = N_{\e}^{(2)}(m_1)\;(1 - 2 \e \log(x) + 2 \e^2 \log(x)^2) \; +\ope(\e^3).
\ee
\end{enumerate}

The routine \ty{ScalIntTwo} performs the integration in all these cases:
\begin{enumerate}
 \renewcommand{\labelenumi}{\alph{enumi})}
 \item ($m_1=m_2$) and b) ($m_2=0$):\\ The propagator structures is replaced with the right hand side of eqs.~(\ref{page55})-(\ref{page68}) respectively:
\be
\label{eq:scalint1}
\mathtt{G[i[m_1,n_1],i[m_2,n_2],i[0,n_3]}\to \f{\pi^4}{(m_1^2)^{n_1+n_2+n_3-4} } \mathtt{N2[m_1]} C^{(2)}_{n_1 n_2 n_3},
\ee
where $\ty{N2[m1]}$ corresponds to eq.~(\ref{ne2}) up to second order in
\ty{eps}
and $C^{(2)}_{n_1 n_2 n_3}$ is given by eq.~(\ref{page55}) in case a) and by eq.~(\ref{page68}) in case b).\item[c)] The integrals are first reduced to the masterintegral, which can then be automatically integrated:
\begin{align}
\mathtt{G[i[m_1,n_1],i[m_2,n_2],i[0,n_3]]} &\rightarrow \mathrm{prefac} (\mathtt{n_1, n_2})\;\mathtt{G[i[m_1,1],i[m_2,1],i[0,1]]} \nnl
&\rightarrow \mathrm{prefac} (\mathtt{n_1, n_2})\,\pi^4 \;m_1^2\,\mathtt{N2} \mathtt{[m_1]}\,\mathtt{C_{111}^{;(2)}},
\end{align}
where the prefactor $\mathrm{prefac} (\mathtt{n_1, n_2})$ depends on the powers $n_1$ and $n_2$ and $\mathtt{C_{111}^{(2)}}
$ is given by eq.~(\ref{masterintegral}).

\item[d)] Factorising two-loop integrals can be directly integrated by making the replacement
\be
\label{eq:scalint2}
\mathtt{AD[i[m_1,n_1],i[m_2,n_2]]}\to 
\f{\pi^4 \;\mathtt{Ne[m_{1}] Ne[m_{2}]} } {(m_1^2)^ {n_1+n_2-4}}C^{(1)}_{n_1} C^{(1)}_{n_2},
\ee
where $C^{(1)}_{n_1} C^{(1)}_{n_2}$ is given by eq.~(\ref{facintc}). 
The replacement rules \ty{nerules} will express the product $\mathtt{Ne[m_{1}] Ne[m_{2}]}$ in terms proportional to $\mathtt{N2[m1]}$ according to eqs.~(\ref{ne11}-\ref{ne22}), if $\mathtt{m_{2}}$ is replaced by $\sqrt {\mathtt{x1}}\; \mathtt{m_{1}}$. Note that in the package \ty{x1} (not \ty{x}) denotes the mass relation $m_{2}^{2}/m_{1}^{2}$.
\end{enumerate}
All the the results of \ty{ScalIntTwo} are expanded in \ty{eps} up to zeroth order.

\section{Documentation of \Fermions}
\label{subsec:docfermions}
\subsection{Declarations}
\label{sub:declarationsfermions}
\begin{description}
\item \ty{DeclareMass[MT,MW,$\ldots$]} used to declare all appearing masses.
\item \ty{DeclareMomentum[q1,q2,k, $\ldots$]} used to declare all appearing momenta.
\item \ty{DeclareIndex[mu,nu,rho,sigma,$ \ldots$]} used to declare all appearing\\ Lorentz indices.
\item \ty{DeclarePolarizationVector[epsilon]} used to declare polarisation vectors, which are treated, except for their properties under
conjugation, in the same way as momenta.
\item \ty{DeclarePolarizationVector[{epsilon,k}]} additionaly sets $epsilon(k) \cdot k =0$.
\end{description}
All of these functions
can be called with an arbitrary number of arguments.
When using one of the newer \Mathematica~
front-ends, it is also possible to use indices in Greek letters
like $\mu$ instead of \ty{mu}.

 \subsection{Dirac Algebra}
\begin{description}
\item \ty{DiracLinearity[expr]} expands all sums within \ty{Dirac[]}
 and takes prefactors of \\masses, momenta and indices out of \ty{Dirac[]}.
 It does the same for \ty{Scal[]}.
\item \ty{DiracAlgebra[expr]} performs the standard Dirac algebra according to \\eqs.~(\ref{metrictensor}), (\ref{anticommutation}) and (\ref{commutation}).
\item \verb~ContractIndex[expr,{mu,nu,...}]~ contracts all Lorentz indices given in curly brackets. For longer exressions \ty{Expand} (or \ty{DiracLinearity}) may have to be used first.
\item \ty{ContractAllIndices[expr]} contracts all silent indices. For longer expressions \\
\ty{DiracLinearity} may have to be used first.
\item \ty{DiracSort[expr,reflist]} orders any sufficiently simple
 expression of $\gamma$s in the order specified in \ty{reflist} (a list
 containing all the momenta and indices appearing in \ty{expr}). For longer \ty{expr} \ty{DiracCollect[expr]} has to be used first.
Projectors (\ty{L}, \ty{R} or \ty{Gamma5}) as well as all momenta have to appear in the \ty{reflist}.
\item \verb#UseDiracEquation[{p,mp},expr,{q,mq}]# sorts \ty{expr}
 and uses the Dirac equation for particles (as in $\bar{u}(p)\; \ty{expr}\; u(q)$). For antiparticles the corresponding syntax is \\
 \verb^UseDiracEquation[{p,-mp},expr,{q,-mq}]^([as in $\bar{v}(p) \;
 \ty{expr} \; v(q)$). \\ In analogy
 \verb@UseDiracEquation[{p,mp},expr,{}]@ and \\
 \verb+UseDiracEquation[{},exp,{q,mq}]+ can be used.
\item \ty{DiracScalExpand[expr]} expands all arguments in \ty{Dirac[]}
 and \ty{Scal[]} (this may be needed in order to contract indices).
\item \ty{DiracCollect[expr]} collects all different Dirac structures.
\item \ty {DiracFactor[expr]} functions like \ty{DiracCollect[expr]} and additionally factorises the coefficient of each of
these different structures.
\end{description}

\subsection{Squaring and Traces}
\label{sec:traces}
The functions presented above are useful at the amplitude level of
a high-energy calculation. In order to obtain physical quantities as
cross-sections and decay-rates, it is necessary to have the tools to conjugate
or square the expressions and to compute traces over products of $\gamma$
matrices. This functionality is provided by the following commands:
\begin{description}
\item \ty{DiracAdjunction} computes the Dirac adjoint ($\bar M = \gamma^0
 M^\dagger \gamma^0$) of a product of $\gamma$ matrices.
\item \ty{DiracSquare[expr,one,two]} returns the trace of\\ $\mathtt{expr \cdot
 one \cdot DiracAdjunction[expr] \cdot two}$, where \ty{one} and \ty{two}
 have to be \\ \ty{Dirac[]} expressions.
\item \ty{DiracTrace[Dirac[...]]} represents the trace over the Dirac
 expression; the trace is not evaluated. Non-Dirac expressions may taken out
 of \\\ty{DiracTrace[...]} using \ty{DiracTraceLinearity}. To evaluate
 the trace,\\ \ty{DiracTraceAlgebra} has to be applied to the expression.
 \item \ty{LearnDiracTraceRule[Dirac[k1,k2,...,kn]]} increases the speed
 of calculations of traces with projectors or with many
 momenta by adding rules to \\
\ty{ExtendedDiracTraceList}. Only momenta, indices or projectors are allowed as input to this
 command. If e.g. \ty{Dirac[k1,...k10]} is entered the routine will also
 learn the rule for any shorter expression of this form (e.g.
 \ty{Dirac[k1,...k8]} and \ty{Dirac[k1,...,k6]}). No rules for the scalar products of these momenta, such as
 \ty{Scal[k1,k2] = 0} are allowed to be implemented.
\end{description}

Traces are evaluated strictly in $D = 4$. For traces over long products of
$\gamma$ matrices it is highly recommended to use \ty{LearnDiracTraceRule}
first in order to significantly speed up the calculation.

Traces involving $\gamma_5$ (and $L$ or $R$) will generally produce terms
involving the $\varepsilon$-tensor (the Levi-Civit\`a symbol). The functions
handling this object are:
\begin{description}
\item \ty{Epsilon[a,b,c,d]} is the completely antisymmetric tensor in four dimensions. The convention $\epsilon^{0123}=-1$ is applied.
\item \ty{EpsilonSort[expr,reflist]} sorts expressions in
 \ty{Epsilon[...]} according to \ty{reflist}.
\item \ty{ContractScalEps[expr]} contracts expressions like\\ \ty{Epsilon[a,...] Scal[a,...]}.
\item \ty{EpsilonEpsilonContract} contracts products of the form\\
 \ty{Epsilon[a,b,c,d]\\ Epsilon[a,e,f,g]}. The same indices
 have to be the first in the list, otherwise\\ \ty{EpsilonSort} has to be used first.

\end{description}

\subsection{Setting Scalar Products On Shell and Replacing of Scalar Products}

In the calculation of physical high energy quantities, momenta are often
restricted by the requirement that particles are on their mass
shell. Furthermore four-momentum conservation allows to express scalar products by
other scalar products thus reducing the number of different
terms. The functions tailored for these needs are:
\begin{description}
\item \verb+SetOnShell[{p1, m1}, {p2, m2}]+ sets $p1 \cdot p1 = m1^2 $ and
 $p2 \cdot p2 = m2^2 $.
 \item \ty{ReplScal[Scal[p1,p2], p1+p2+q1==q2]} generates a
 replacement list of the form\\ $p1 \cdot p2 \rightarrow \f {1}{2}
(-(p1^2) - 2 p1 \cdot q1 - p2^2 - 2 p2 \cdot q1 -
2 q1 \cdot q1 + q2^2)$ obtained by squaring both sides of the identity given as second argument of the function. So \ty{q2==p1+p2+q1} will produce a different result than \ty{-q1==p1+p2-q2}. \ty{-q-p1==p2-p} will return an empty list.
\end{description}

\section{Documentation of \Integrals}
\label{subsec:docintegrals}
\subsection{Additional Declarations}
\label{sub:declarationsintegrals}
\begin{description}
\item \ty{DeclareSmallMass[MU, MD]}: Needed for \ty{Scaling}.
\item \ty{DeclareHeavyMass[MT,MW]}: Needed for \ty{TaylorMass}.
\item \ty{DeclareExternalMomentum[k1,k2]}: Needed for \ty{Scaling} and\\ \ty{TaylorExpansion}.
\item \ty{DeclareLoopMomentum[q1,q2]}: Needed for \ty{TaylorExpansion}.
\end{description}

\subsection{Transformation of the Integrals to Scalar Integrals}
\bit
\item \ty{Color} replaces colour structures depending on generators and structure constants of $SU(3)_c$ on expressions only depending on generators or scalars corresponding to \\eqs.~(\ref{c1}-\ref{c2}).
\item \ty{TaylorExpansion} expands denominators of the form\\
\ty{AD[\ldots, den[q+k, m],\ldots ]}, where \ty{q} is a loop momentum or the sum of loop momenta, in k up to second order. Note that loop momenta have to be declared with \ty{DeclareLoopMomentum} first.
\item \ty{TaylorMass} expands denominators of the form
\ty{AD[\ldots, den[q, m], \ldots]},\\where \ty{q} is a loop momentum, in \ty{m} up to second order, if m is NOT declared as heavy mass with\ty{ DeclareHeavyMass}.
\item \ty{Scaling} multiplies all momenta declared as external with \ty{DeclareExternalMomentum} and all masses declared as small with \ty{DeclareSmallMass} with a factor $x$ and sets then all powers $ x^n$ with $ n>2$ a to zero.
\item \ty{PartialFractionOne} makes partial fraction of the denominators in the one-loop case according to eqs.~(\ref {eq:partial_one}-\ref{simplification}) and successively gets rid of loop momenta from the numerators successively according to eq.~(\ref{red1}).
\item \ty {PartialFractionTwo} makes partial fraction of the denominators in the two-loop case according to eqs.~(\ref{eq:partial_one}-\ref{simplification}) and successively gets rid of loop momenta from the numerator according to eqs.~(\ref{red1})-(\ref{red2}) and sets all vanishing massless integrals to zero.
\item \ty{TensorOne[expr,var]} performs the one-dimensional tensor reduction in \\\ty{var}.
It assumes that the denominator of \ty{expr} is an arbitrary scalar function depending on
Lorentz invariants of var. It can handle expressions \ty{expr} with up to 9 Lorentz Indices. Results are Taylor expanded in \ty{eps} up to second order.

\item \ty{TensorTwo[expr,var,var2]} performs a two dimensional tensor reduction of expressions \ty{expr} with up to five Lorentz Indices
assuming that the denominator of \ty{expr} is an arbitrary scalar function of the variables \ty{var1} and \ty{var2}. If the numerator of \ty{expr} depends only on \ty{ var} (\ty{var2}), it performs a one-dimensional tensor reduction in \ty{var} (\ty{var2}) using
\ty{TensorOne[expr,var]} (\ty{TensorOne[expr,var2]}). Expressions like \ty{Scal[var,var]} are treated like \ty{Scal[var,lor1]Scal[var,lor1]}, where \ty{lor1} is a Lorentz index. This artificially increases the number of used Lorentz indices. Therefore it is recommended to set all quadratic scalar products to a dummy variable before performing the tensor reduction. Results are expanded in \ty{eps} up to second order. Factorising two-loop integrals have to be tensor-reduced with \ty{TensorOne} before usage of \ty{TensorTwo}.

\item \ty{SimplifyPropagator} brings propagator structures to the form needed for loop integration.
\eit

\subsection{Integration of Scalar Integrals}
\bit
\item
\ty {ScalIntOne[expr]} allows the calculation of scalar one-loop integrals by replacing the propagator structure \ty{AD[i[m,n]]} by the right hand side of eq.~(\ref{intoneloop}):
 \be
 \mathtt{AD[i[m,n]]} \rightarrow\ \f{\pi^2}{ (m^2)^ {n-2}} \mathtt{Ne(m)\; C^{(1)}_n},
 \ee
where \ty{Ne(m)} corresponds to eq.~(\ref{Ne}) and $\mathtt{C^{(1)}_n}$ to eq.~(\ref{oneloop}) up to second order in \ty{eps}. Results are expanded up to first order in \ty{eps}.
\item
\ty{ScalIntTwo[expr]} allows the calculation of scalar integrals independent of external momenta and with one massless propagator. It replaces in \ty{expr} propagator structures of the form
$
\mathtt{
G[i[m1, n1], i[m2, n2], i[0, n3]]}
$
with the analytical result of the corresponding scalar twoloop integral
$G_{n1,n2,n3}^{m1,m2,0}$ as defined in eqs.~(\ref{eq:scalint1}-\ref{eq:scalint2}). Note that the result is expanded in \ty{eps} up to zeroth order.
\item
\ty {ScalIntTwoThreeMasses[expr]} allows the calculation of scalar loop integrals independent of external momenta and with up two different masses expressions of the form
$
\mathtt{G[i[m1, n1], i[m2, n2], i[m1, n3]]}
$
with the analytical result for scalar two-loop integrals of the form
$G_{n1,n2,n3}^{m1,m2,m1}$ as defined in eq.~(\ref{masteri}). Note that the result is expanded in \ty{eps} up to zeroth order.
\item 
\ty{nerules} are replacement rules allowing to express prefactors \ty{Ne[M1]Ne[M2]} in terms proportional to \ty{Ne[M2]}, if\ty{M2} is given as \ty{M2=Sqrt[x1]*M1}.
\eit

\section{Installation Instructions}
\label{sec:installation}
The package \MasterTwo\ can be downloaded from \\ 
\ty{https://github.com/shhschilling/MasterTwo}
\subsection{Installation under Linux}

Copy the zip file MasterTwo-1.0.zip to your disk and unpack it with

\ty{> gunzip MasterTwo-1.0.zip}

Change to the MasterTwo-1.0 directory

\ty{> cd MasterTwo-1.0}

and change the permission of the installation script:

\ty{> chmod +x MasterTwoInstall}

Execute it with

\ty{> ./MasterTwoInstall}

and follow the instructions. The installation package will update the init.m file in the .Mathematica/Autoload/ directory, so that you can load the package without having to give the whole path.

\subsection*{Uninstallation under Linux}

Run the program

\ty{> ./MasterTwoUninstall}

in the installation directory of MasterTwo.

\subsection{Installation under Windows and MacOs}
 \MasterTwo\ must be copied to one of the \Mathematica~ Autoload paths. 
\bit
\item 
Type 
\ty{\$Path}
 on the \Mathematica~ command line to identify the Autoload paths on your system.
 \item 
 Copy the files \ty{Fermions.m, Integrals.m} and \ty{MasterTwo.m} in one of the Autoload paths of your Mathematica installation.
 \item Close Mathematica.
 \item In a new \Mathematica~ session, call the package package \MasterTwo\ by typing
\begin{description}
 \item\ty{<<MasterTwo\`} 
 \end{description}
on the command line.
\eit
From here on all the \MasterTwo\ commands are available.
The package is equipped with an on-line help to each
command. \verb+MasterTwoInfo[]+ produces a list with all the available
commands and \ty{?command} prints a short information on syntax and effect
of \ty{command}.

\section{Generation of the Integrands: \FeynArts~and \MasterTwo}
\label{sec:FeynArts}
	The natural starting point for the generation of the integrands of the one and two-loop integrals to be integrated is the usage of the programme \FeynArts~\cite{FeynArts3}.
The existing model files for the SM-model and the SUSY extensions can be easily adapted to the conventions needed for the processes to be calculated.
But the Feynman amplitudes generated by \FeynArts\ are not appropriate for the routines used in \MasterTwo. Thus the function \ty{FeynArtsToMasterTwo} translates Standard Model output generated by \FeynArts\ into a form adapted for the further usage of \MasterTwo. In the following we list the most important automatic replacements:
 \bit
\item \textit{Renaming of the headers}
\begin{align}
\mathtt{FeynAmp[GraphName[...],Integral[...],c]}&\rightarrow \mathtt{c},\nnl
\mathtt{FermionChain} &\rightarrow \mathtt{Dirac},\nnl
\mathtt{PropagatorDenominator} &\rightarrow \mathtt{den},\nnl
\mathtt{MetricTensor}& \rightarrow \mathtt{Scal},\nnl
\mathtt{FourVector}&\rightarrow \mathtt{Scal}\nonumber
\end{align}
Note that in the first replacement only the integrand of the integral is kept. Thus in the final calculation of scalar integrals we will actually replace the propagator structure of scalar integrands with the value of the corresponding scalar integral. 
\item \textit{Replacements in Fermion chains}
\begin{align}
\mathtt{ChiralityProjector[-1]}&\rightarrow \mathtt{L}, \nnl
\mathtt{ChiralityProjector[1]}&\rightarrow \mathtt{R},\nnl
\mathtt{DiracSlash[a]}&\rightarrow \mathtt{a}\nonumber
\end{align}
 \item \textit{Renaming of the momenta}\\
 \FeynArts~ declares internal momenta by
 \ty{FourMomentum[Internal, i]}, external momenta as
  \ty{FourMomentum[External, j]}, where $\mathtt{i}=1,...,\mathtt{l}$ ($\mathtt{j}=1,...,\mathtt{k}$) stands for the \ty{i.} (\ty{j.}) internal (external) momenta appearing in the diagram. \ty{FeynArtsToMasterTwo} makes then the following replacements:
  \begin{align}
 \mathtt{FourMomentum[Internal, j]} &\rightarrow \mathtt{qj},\nonumber \\
 \mathtt{FourMomentum[Outgoing, j]} &\rightarrow \mathtt{kj} \nonumber
 \end{align}
 \item \textit{Renaming of Lorentz indices}
\be
 \mathtt{DiracMatrix[Index[Lorentz,a]]}\rightarrow \mathtt{lora}\nonumber
 \ee
\item{\textit{Dirac Spinors}}\\
Dirac Spinors are by default set to one by making the replacement
\be
\mathtt{DiracSpinor[a_{\_}]}\rightarrow\mathtt{Times[]} \nonumber
\ee
If the user wants to use the Dirac equation or is interested in the calculation of squared matrix elements etc. this replacement should be commented.
\eit

The function \ty{FeynArtsToMasterTwo} depends very much on the concrete process to be calculated and has to be adapted when using new model files, calculating different processes, using newer versions of \ty {FeynArts} etc.

\section{Example of a Two Loop Diagram}
\label{sec:example}
\begin{figure}[htb]
\begin{center}	
\includegraphics[width=8cm]{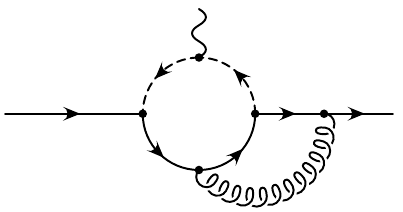}	
\end{center}	
\caption
{\small
 {Example: A one-particle irreducible two-loop diagram for 
$b \to s \gamma$ . The external quark lines (solid) denote the incoming $b$-quark and the
outgoing $s$-quark, respectively. The wavy line denotes a virtual photon. The internal dashed-, solid-
and curly lines denote the charged W boson $W^\pm$, the $t$-quark and the
gluon, respectively.}} 
\label{fig:feyndiag2}
\end{figure}

In Figure .~\ref{fig:feyndiag2} we show an example diagram of the two-loop decay $\bsgamma$. Its calculation with the help of \MasterTwo\ is given in the file \ty{Example.nb} included in this distribution. 

\section{Acknowledgements}
S. Schilling would like to thank D. Wyler for helpful discussions. Special thanks to M. Misiak and J. Urban for fruitful discussions and
advice regarding the technical details of the two-loop calculations, P. Liniger for the provision of the original code and documentation of \Fermions\ as well as K. Bieri for some routines concerning the tensor reduction now implemented in \Integrals. This work was partially supported by the Swiss National Foundation.


\end{document}